\documentclass[twocolumn, amssymb, amsmath, nobibnotes, nofootinbib, aps, prb,showpacs]{revtex4}
\usepackage[dvips]{graphicx}
\begin{document}
\title{Multiple first order transitions and associated room temperature magneto-functionality in Ni$_{2.048}$Mn$_{1.312}$In$_{0.64}$}
\author{S. Pramanick$^1$}
\author{P. Dutta$^2$}
\author{S. Chatterjee$^2$}
\author{S. Giri$^1$}
\author{S. Majumdar$^1$}
\email{sspsm2@iacs.res.in} 
\affiliation{$^1$Department of Solid State Physics, Indian Association for the Cultivation of Science, 2A \& B Raja S. C. Mullick Road, Jadavpur, Kolkata 700 032, INDIA}
\affiliation{$^2$UGC-DAE Consortium for Scientific Research, Kolkata Centre, Sector III, LB-8, Salt Lake, Kolkata 700 098, INDIA}


\begin{abstract}
Present work reports on the observation of multiple magnetic transitions in  a Ni-excess ferromagnetic shape memory alloy with nominal composition Ni$_{2.048}$Mn$_{1.312}$In$_{0.64}$. The magnetization data reveal two distinct thermal hystereses associated with two different phase transitions at different temperature regions. The high temperature magnetic hysteresis is due to the martensitic phase transition whereas the low temperature hysteresis  occurs around the magnetic anomaly signifying  the transition from a paramagnetic-like state to the ferromagetic ground state within the martensite. Clear thermal hysteresis along with the sign of the curvatures of Arrott plot curves confirm the {\it first order nature of both the transitions}. In addition, the studied alloy is found to be functionally rich with the observation of large magnetoresistance (-45\% and -4\% at 80 kOe) and magnetocaloric effect (+16.7 J.kg$^{-1}$.K$^{-1}$ and -2.25 J.kg$^{-1}$.K$^{-1}$ at 50 kOe) around these two  hysteresis regions (300 K and 195 K respectively).   
\end{abstract}
\maketitle

\section{Introduction}
In recent times Ni$_2$Mn$_{1+p}$Z$_{1-p}$ (Z= In, Sn, Sb, and $p <$ 1) based ferromagnetic shape memory alloys (FSMAs) emerge out to be  novel functional materials alongside with the manifestation of intriguing physical properties~\cite{krenke,kainuma,manosa,khan}. Such novelty in these alloys can aptly be attributed to the existence of mutual correlation between structural, magnetic and electronic degrees of freedom. The alloys are characterized by first order martensitic phase transition (MPT) at $T_{MS}$ as well as a second order paramagnetic (PM) to ferromagnetic (FM) transition at the Curie point ($T_C$)~\cite{sutou,koyama,souvik1}. For the realization of a magneto-functional alloy at around $T_{MS}$, it is essential that $T_C > T_{MS}$, {\it i.e.}, the sample should be in a magnetically ordered state when MPT takes place. MPT also tends to alter the magnetic  state of the alloy and it has been reported that sample may lose its ferromagnetism or can attain a state having antiferromagnetic (AFM)-like correlation below $T_{MS}$~\cite{enkovara,brown,aksoy}. The situation can be rather exotic when second order  $T_C $ and first order $ T_{MS}$ lie close to each other in a particular alloy~\cite{Enric}. Notably, the functional properties such as magneto-caloric effect (MCE) in a sample depends strongly on the order of the phase transition.

\par
In order to address such issues in a Ni-Mn-Z based FSMA, we investigated a Z =  In alloy with nominal composition Ni$_{2.048}$Mn$_{1.312}$In$_{0.64}$. Here the Ni stoichiometry is greater than 2 and this calculated amount of excess Ni is doped at the expense of Mn to enhance $T_{MS}$ and to bring it close to $T_C$. Among  various members of Ni-Mn-Z family, Z $=$ In  alloys are  known to have the largest functionalities~\cite{krenke3,sharma,Yu,sabya1}. For the studied alloy both $T_C$ and $T_{MS}$ are close  to 300 K which provides an added advantage for practical applications. $T_{MS}$ being comparatively large, the sample loses its ferromagnetically ordered state in the martensite, although it recovers the same at a lower temperature. Consequently, the sample has fascinating phase diagram with two magnetic transitions, one in the martensite and the other in the austenite (we denote them by $T_{CM}$ and $T_{CA}$ respectively) along with  one structural transition very close to $T_{CA}$~\cite{krenke4,dls,wang,sabya2}. The present work is devoted to understand multiple phase transitions in the alloy as well as to study its magneto-functional properties and it is based upon electrical transport and magnetic measurements on the alloy. 

\section{Experimental Details}
Polycrystalline sample of nominal composition Ni$_{2.048}$Mn$_{1.312}$In$_{0.64}$ was prepared by arc melting technique. For homogeneity, the sample was remelted several times turning the ingot back to back. Next, the ingot was vacuum sealed in a quartz capsule and annealed at 900$^{\circ}$ C for 48~h followed by a rapid ice water quenching~\cite{krenke,kainuma,sutou}.  Resistivity ($\rho$) was measured by four probe method on a homemade setup fitted in a nitrogen cryostat as well as on cryogen-free high magnetic field system (Cryogenic Ltd., U.K.). The magnetic measurements were performed by using a commercial Quantum Design SQUID magnetometer (MPMS XL Ever Cool model).
\begin{figure}[t]
\centering
\includegraphics[width = 8 cm]{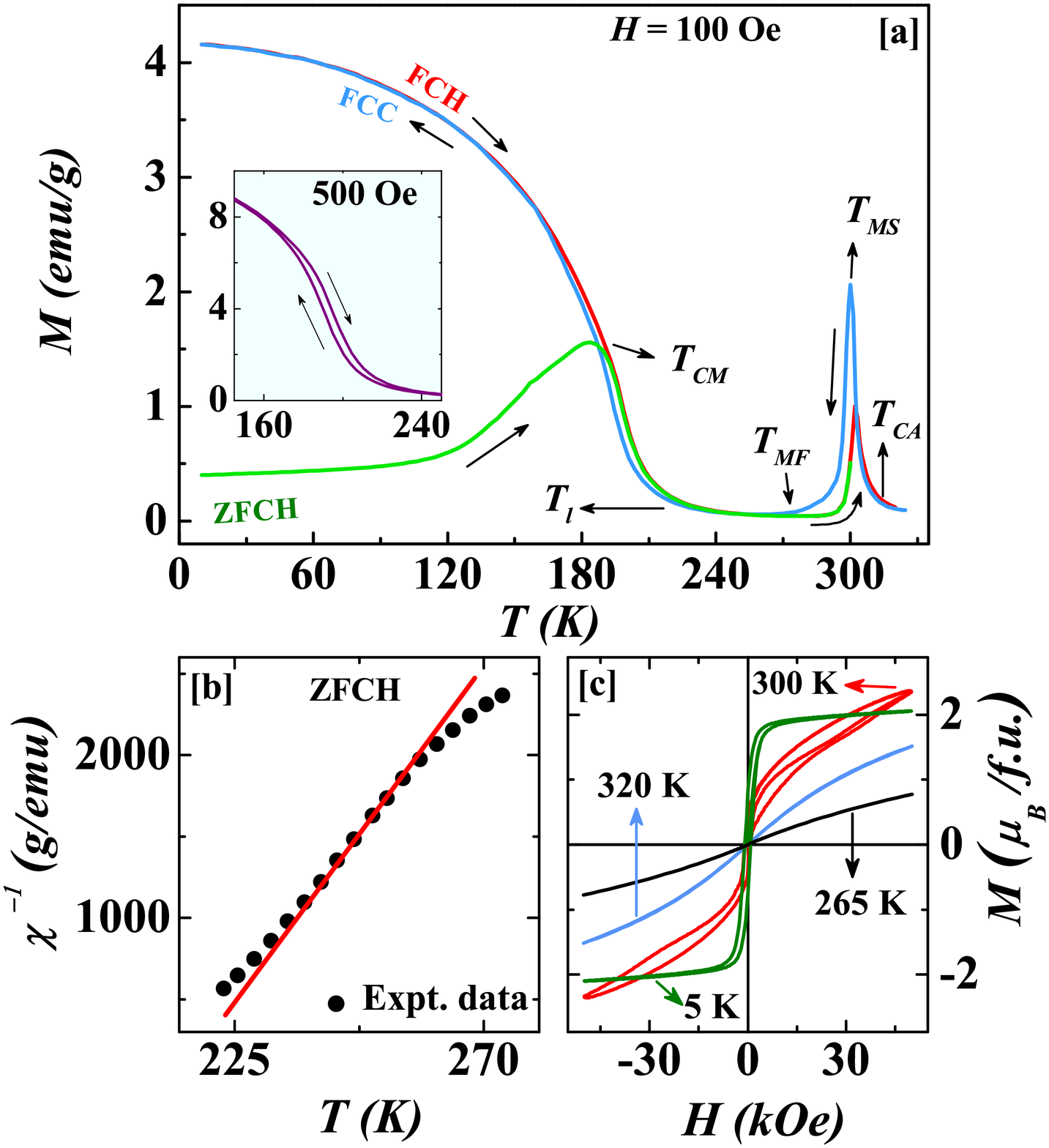}
\caption {Main panel of (a)  depicts the thermal variation of $M$ measured in 100 Oe. The inset shows $M$-$T$ data at 500 Oe measured in FCC and FCH modes between 150-250 K depicting thermal hysteresis. (b) shows the inverse susceptibility {\it vs.} temperature data between 222-273 K. The solid line is a guide to the eye. (c) shows isothermal $M$-$H$ curves at 5 K, 265 K, 300 K and 320 K respectively for field cycling between $\pm$ 50 kOe.}
\end{figure}

\section{Results and Discussions}
Main panel of the fig. 1 (a) depicts temperature ($T$) variation of  magnetization ($M$) for the present sample measured by the following three protocols, namely (i) zero-field-cooled-heating (ZFCH), (ii) field-cooled-cooling (FCC) and (iii) field-cooled-heating (FCH) in  presence of 100 Oe of applied magnetic field ($H$). The $M-T$ data show multiple features depending upon the change in magnetic phases due to thermal variation. At around 310 K, $M(T)$ shows a sharp upturn with decreasing $T$, which corresponds to the PM to FM transition point in the austenite ($T_{CA}$). On further cooling  $M$ attains a peak ($T_{MS}$) followed by a sharp fall  which is also associated with thermal hysteresis between $M_{FCH}(T)$ and $M_{FCC}(T)$ data between 300-265 K.  Such hysteresis indicates first order transition and we identify it to be associated with the  MPT. The fall in $M$ below $T_{MS}$  is possibly due to the loss of ferromagnetism and the sample presumably becomes PM below the MPT~\cite{moss1,moss2}. For further decrease in $T$ below 265 K, $M$ remains almost unchanged till 225 K (denoted as $T_{l}$) in the FCC and FCH measurements. However, it regains its ferromagnetism on further cooling and starts to rise again below about 225 K and shows a peak (in the ZFCH data only) around 185 K which  can be considered as the Curie point of the martensite ($T_{CM}$). Interestingly, a second thermal hysteresis loop is formed between FCC and FCH data in the region 225-165 K, which is just around $T_{CM}$. This hysteresis is visible even in higher $H$ and an enlarged view of the heating and cooling data measured at $H$ = 500 Oe are shown in the inset of fig. 1 (a).  The ZFCH data branch out from FCH right from 190 K. Below 185 K, ZFCH shows a  fall and it monotonically decreases down to the lowest $T$ of measurement. Such thermomagnetic irreversibility is quite common and observed previously in many other Ni-Mn-Z based  shape memory alloys~\cite{souvik2,sabya3}. Possible origin  of  the observed irreversibility may be the  pinning of FM domains by the martensitic variants and/or by the phase boundaries.
\par
We analysed the magnetic susceptibility ($\chi$ = $M/H$) data of the PM-like phase in the martensite (between $T_{MF}$ and $T_l$) by plotting $\chi^{-1}$ {\it vs.} $T$ [Fig. 1 (b)]  measured in ZFCH mode. The curve does not show a linear trend expected for an ideal PM system. Such deviation from linearity can be attributed to the phase coexistence in the vicinity of MPT.    
\par
Multiple feature enriched $M$-$T$ data provoked us to measure the isothermal field variation of $M$ at different temperatures to unveil the true nature of the magnetic states. Fig. 1 (c) shows the measured isotherms in the  ZFC state at 5 K, 265 K, 300 K and 320 K for field cycling between $\pm$ 50 kOe. The 5 K isotherm shows FM nature with almost saturated magnetization value of 2.04 $\mu_B$/f.u. On the other hand, PM-like nature is obtained from the 320 K isotherm as there is no tendency of saturation of $M$ up to $H$ = 50 kOe. However, isotherm at 320 K shows finite curvature which  possibly arises due to the existence of short range correlations prior to long range order. The isotherm within the region of MPT shows quite different nature as evident  from the 300 K data. The sample shows virgin line effect with initial leg of field ramping remaining outside the subsequent field cycling curves. Apart from the virgin line effect, distinct loop is observed on field cycling beyond the virgin curve. Such loops are formed away from the origin in the first and third quadrants of the $M$-$H$ coordinates. The nature of the curve points towards a partial FM state as complete saturation of $M$ is absent. For the studied sample $T_{CA}$ and $T_{MS}$ are almost overlapping and consequently the  state around 300 K is a mixed one with combined FM (austenite)  and PM (martensite)  phase fractions, resulting the non-saturated $M$-$H$ isotherm. Interestingly, the 265 K isotherm shows completely different nature from an FM  state although it is below  $T_{CA}$. The behavior  rather resembles to 320 K one with a lower value of $M_{50kOe}$(=0.77 $\mu_B/$f.u.). This is due to the fact the at 265 K the sample loses its long range FM order due to the occurrence of MPT and goes into a PM-like state. It is already known that martensite with tetragonal or orthorhombic crystal symmetry has lower FM $T_C$ than the cubic austenite~\cite{monte}. This PM-like state approximately extends between  $T_{MF}$ and $T_{l}$ on the $M$ {\it vs.} $T$ curve (FCC) in Fig. 1 (a). The moment in the PM-like state below $T_{MS}$ is lower than the PM-state above $T_{CA}$ due to the existence of intersite AFM correlation between Mn sitting at In site and regular Mn atoms~\cite{monte,xps,tan}. 
\begin{figure}[t]
\centering
\includegraphics[width = 8 cm]{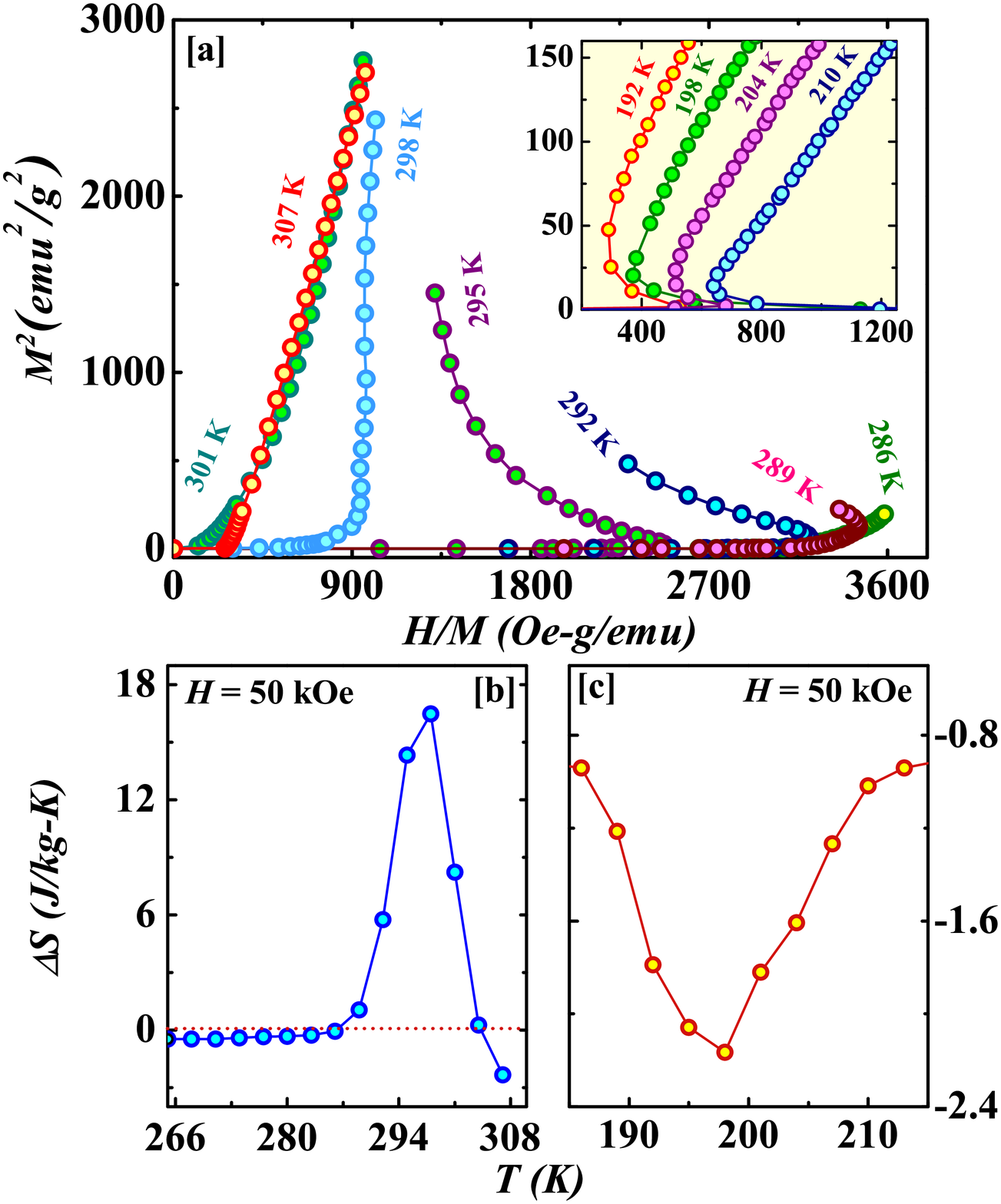}
\caption {(a) Main panel depicts $M^2$ {\it vs.} $H/M$ plot around MPT temperature. Inset shows the same around $T_{CM}$. (b) and (c) shows field-induced entropy change as a function of the temperature measured around MPT and $T_{CM}$ respectively for the applied field of 50 kOe.}
\end{figure}
\par
In order to confirm the order of the transitions, we performed Arrott plot analysis of the isothermal $M$ {\it vs.} $H$ data as shown in fig. 2 (a). We have constructed the Arrott plots ($M^2$ {\it vs.} $H/M$ curves) across the transitions $T_{MS}-T_{CA}$ (main panel of fig. 2 (a))  and around $T_{CM}$ (inset of fig. 2 (a)). It is known that for a first order phase transition (FOPT),  Arrott plot is associated with negative curvature~\cite {art1,art2,art3,art4,art5}. In the $T_{MS}-T_{CA}$ regime, the Arrott plots above 300 K show positive curvature at par with second order PM to FM transition. However, below 300 K, signature of negative curvature is evident, which is associated with the first order MPT occurring below 300 K~\cite {crit1,crit2}. The second magnetic transition occurring around $T_{CM}$ is also probed by Arrott plot technique and it  shows the existence of negative curvature. This is in line with  the observed thermal hysteresis around that transition in the $M$ {\it vs.} $T$ data.   

\par
Our study indicates that the low-$T$ FM transition occurring in  martensite  is first order in nature, while the transition close to room temperature has mixed character depending upon $T$ due to close proximity of $T_{CA}$ and $T_{MS}$. Recently, it has been shown that functional properties such as MCE can be maximum in the FSMA when $T_{CA}$ and $T_{MS}$ coincide~\cite{Enric}. Considering the closeness of  these temperatures in the present alloy, we measured MCE in terms of change in entropy ($\Delta S$) due to the application of $H$  around $T_{MS}-T_{CA}$ regime. Also, we investigated $\Delta S$ around the first order magnetic transition at $T_{CM}$.  $\Delta S$ can be calculated from the isothermal $M(H)$ data by using Maxwell's thermodynamical relation $$\Delta{S}(0\rightarrow H_0) = \int^{H_0}_{0}\left(\frac{\partial{M}}{\partial{T}}\right )_HdH,$$ where $\Delta{S}(0\rightarrow H_0)$ denotes the entropy change for the change in $H$ from 0 to $H_0$. For this purpose isothermal $M(H)$ data at different temperatures were recorded in 3 K interval around the MPT as well as around $T_{CM}$. Thermal variation of calculated $\Delta S$ for $H$ changing from 0 to 50 kOe in the vicinity of  $T_{MS}$ and $T_{CM}$ are plotted in fig. 2 (b) and (c) respectively. $\Delta S$ is found to be positive around $T_{MS}$ having the peak value of about 16.7 J.kg$^{-1}$.K$^{-1}$ (for $H_0$ = 50 kOe). The positive value of $\Delta S$ indicates the presence of inverse MCE as reported previously in other Ni-Mn-Z FSMAs. Now coming to the second magnetic transition, $\Delta S$ shows a peak structure around $T_{CM}$. Here $\Delta S$ is conventional ({\it i.e.} negative) having the peak value of about  2.25 J.kg$^{-1}$.K$^{-1}$ (for $H_0$ = 50 kOe) at 195 K.

\begin{figure}[t]
\centering
\includegraphics[width = 8 cm]{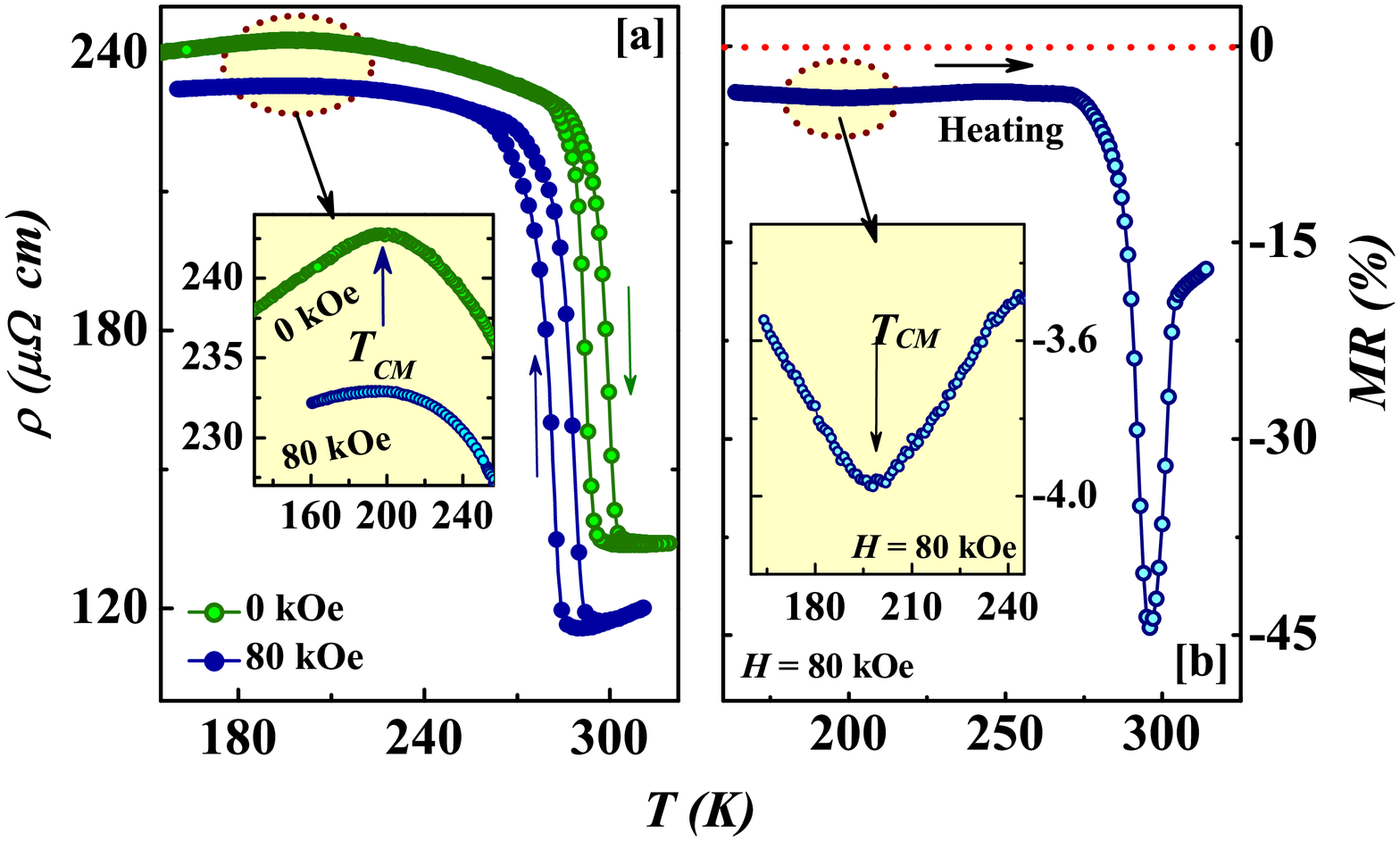}
\caption {(a) Main part shows thermal variation of $\rho$ (both during heating and cooling) in presence of applied field $H$ =0 and 80 kOe. Inset depicts zoomed view of $\rho$ - $T$ around $T_{CM}$. (b) Shows $T$ variation of MR calculated from $\rho (T)$ data. Enlarged view of MR around $T_{CM}$ is shown in the inset of (b).}
\end{figure}

\par
We also probed these transitions via magneto-transport measurements as shown in fig. 3. $T$ variation of $\rho$ in zero field as well as in  $H$ =  80 kOe for both the heating and subsequent cooling protocols are presented in the main panel of Fig. 3(a). The onset of MPT in the form of sharp jump in $\rho$ during cooling is observed around $T_{MS}$ = 299 K and the transition is marked by thermal hysteresis.  $T_{MS}$ shifts to lower $T$  in presence of  $H$ with a rate of 2 K/ 10 kOe.  We have calculated the MR using the formula $\left(= \left[\frac{{\rho(H)-\rho(0)}}{\rho(0)}\right] \right)$ from the heating path of $\rho$ vs $T$ data and plotted in  fig. 3(b). Maximum value of MR of about -45\% is obtained for $H$ = 80 kOe  around 300 K. The significant contribution of negative MR is mostly concentrated around thermal hysteresis region associated with the MPT (roughly between  280-300 K), pointing towards the field sensitivity of the metastable phases around the region.

\par
Zero field resistivity shows a hump like character around 195 K [see inset of fig.3 (a)], which matches well with the second magnetic transition $T_{CM}$ occurring in the martensite. This hump like region becomes  flatter under the application of $H$ resulting negative MR. $T$ variation of MR around $T_{CM}$ shows a dip with the maximum value of MR being -4\% [inset of fig.3 (b)].

\section{Summary and Conclusions}
The present work reports large MCE and MR at room temperature in the excess Ni doped (compared to the Heusler stoichiometry) alloy Ni$_{2.048}$Mn$_{1.312}$In$_{0.64}$. The salient point is that here the FM curie point and the $T_{MS}$ are very close together and that presumably boosts the functional properties of the alloy. Addition of excess Ni at the expense of Mn does not alter the Curie point much but it enhances the MPT temperature and thereby makes the sample useful for practical applications.

\par
An important outcome of our investigation is the order of phase transitions. Although the PM to FM transition in the austenite above room temperature is found to be second order in nature, surprisingly the low-$T$ FM transition in the martensite at $T_{CM}$ is clearly {\it first order}. It is well known that there is a strong inter-correlation between magnetism and lattice in these alloys~\cite{souvik2}. The FM transition at $T_{CM}$ takes place from a PM state which originates from the loss of ferromagnetism due to the MPT. This PM state (between $T_{l}$ and $T_{MF}$) is having incipient inter-site AFM correlation between Mn sitting at Y and Z sites (refer Heusler formula Ni$_2$YZ) and does not obey a simple Curie-Weiss law. So, this reentrant FM state below $T_{CM}$ is  originating from a rather unconventional PM state and  development of ferromagnetism may lead to structural changes. The observed FOPT at $T_{CM}$ may  indicate some sort of  lattice distortion which paves the path for the FM state to occur. Very recently, it has been reported for a similar alloy (Ni$_2$Mn$_{1.44}$Sb$_{0.32}$Ga$_{0.24}$) that crystal lattice indeed undergoes structural transition  at the second FM transition (this is equivalent to $T_{CM}$ in our symbol) from a 7M orthorhombic to a modulated 4O orthorhombic phase which is either second order or weakly first order~\cite {Tian}. In analogy with this Sb-Ga based alloy, $T_{CM}$ may also be associated with some sort of structural change. At this point it is difficult to comment on the exact nature of such changes.  Possibly the  weak first order transition is magneto-elastic in nature, where the $M$ is the primary order parameter which ultimately leads to a structural distortion.  
\par
In conclusion, we observe large MCE and MR at room temperature in a Ni-excess FSMA of nominal composition Ni$_{2.048}$Mn$_{1.312}$In$_{0.64}$. The alloy interestingly shows multiple magnetic and structural transitions and our study indicates that the low-$T$ FM transition in the martensite is  {\it first order in nature} and it may be associated with some lattice distortion.

\section{Acknowledgment}
The work is supported by the grants from BRNS, India (2012/37P/39/BRNS/1991). Authors would like to thank Department of Science and Technology, India for low temperature high magnetic field facilities at UGC-DAE Consortium for Scientific Research, Kolkata Centre.

{99}

\end{document}